\documentclass[aps,twocolumn,groupedaddress,showpacs]{revtex4}

\usepackage{graphicx}
\usepackage{epsfig}

\begin{document}

\title{How to detect the fourth order cumulant of electrical noise}

\author{Joachim~Ankerhold$^{1,2}$ and Hermann~Grabert$^{1}$}

\affiliation{${}^1$ Physikalisches Institut,
Albert-Ludwigs-Universit\"at,
 79104 Freiburg, Germany\\
${}^2$ Service de Physique de l'Etat Condens\'{e},
DSM/DRECAM, CEA Saclay, 91191 Gif-sur-Yvette, France}

\date{\today}

\begin{abstract}
It is proposed to measure the current noise generated in a
mesoscopic conductor by macroscopic quantum tunneling (MQT)
in a current biased Josephson junction placed in parallel
to the conductor. The theoretical description of this
set-up takes into account the complete dynamics of
detector and noise source. Explicit results are given for
the specific case of current fluctuations in an oxide
layer tunnel junction, and it is shown how the device
allows to extract the fourth order cumulant of the noise
from the MQT data for realistic experimental parameters.
\end{abstract}


\maketitle

Within the last decade electrical noise has moved into the
focus of research activities on electronic transport in
nanostructures \cite{bb}, since it provides information on
microscopic mechanisms of the transport not available from
the voltage dependence of the average current. Lately,
attention has turned from the noise auto-correlation
function to higher order cumulants of the current
fluctuations characterizing non-Gaussian statistics
\cite{lll,yn}. While theoretical attempts to predict these
cumulants for a variety of devices are quite numerous
\cite{yn}, experimental observation is hard because of
small signals, large bandwidth detection, and strict
filtering demands. A first pioneering measurement by
Reulet {\sl et al.} \cite{rsp} of the third cumulant of
the current noise from a tunnel junction has intensified
efforts and several new proposals for experimental set-ups
have been put forward very recently, some of which are
based on Josephson junctions (JJ) as noise detectors.
Lindell {\sl et al.} \cite{lds} employed a Coulomb
blockaded JJ to demonstrate that the conductance of the
junction in the Coulomb gap region is sensitive to the
non-Gaussian character of noise applied to the junction. A
modification of this set-up was suggested by Heikkil\"a
{\em et al.} \cite{heik} to get specific information on
the third cumulant of the noise. Another recent experiment
\cite{pek2} has observed activated-over-the-barrier-jumps
of a JJ biased by a noisy current. The data are consistent
with resonant activation produced by the second cumulant
of the noise at the plasma frequency of the junction. For
a measurement of the full distribution of current
fluctuations Tobiska and Nazarov \cite{tn} suggested to
use an array of overdamped JJs acting as a threshold
detector for rare current fluctuations triggering
over-the-barrier jumps. Since in the overdamped limit
retrapping spoils the built-up of a detectable voltage, it
was argued by Pekola \cite{pek1} that an experimentally
more accessible detector would extract the noise
characteristics from modifications of the macroscopic
quantum tunneling (MQT) rate in an underdamped JJ.

\begin{figure}
\epsfxsize=8cm \epsfysize=4cm \epsffile{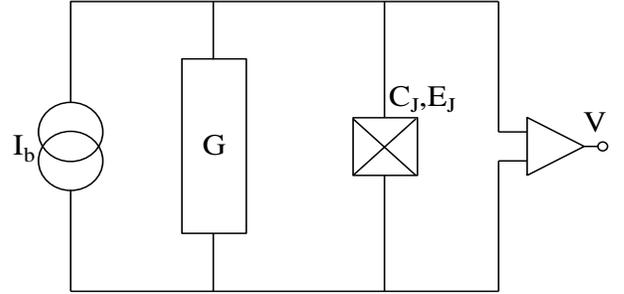}
\caption[]{\label{fig1}Electrical circuit containing a
mesoscopic conductor $G$ in parallel to a JJ with
capacitance $C_J$ and coupling energy $E_J$ biased by an
external current $I_b$. The switching out of the zero
voltage state of the JJ by MQT is detected as a voltage
pulse $V$.}
\end{figure}

In all experimental set-ups to measure higher order
cumulants realized and proposed so far, heating is one of
the major experimental obstacles \cite{reulet}. Thus,
experiments have primarily attempted to establish just the
unspecified non-Gaussian nature of the noise or to measure
the third cumulant (skewness). The latter one is
particularly accessible since it can be discriminated from
purely Gaussian noise due to its asymmetry, e.g.\ when
inverting the current through the conductor. This is in
contrast to the fourth order cumulant (sharpness), which on
the one hand due to heating effects may be completely
hidden behind the second and the third one, but on the
other hand is required to gain an essentially complete
characterization of the distribution of current
fluctuations. In this Letter we propose and analyze a
set-up, with the circuit diagram depicted in
Fig.~\ref{fig1}, which allows to detect the fourth order
cumulant of the current noise generated by a nanoscale
conductor. Since this conductor is placed in parallel to a
current biased JJ in the zero voltage state, no heating
occurs prior to the decay of this state by MQT. However,
the MQT rate is modified in a specific way by the even
higher order cumulants characterizing the non-Gaussian
current fluctuations of the conductor.

The complete statistics of current noise generated by a
mesoscopic conductor can be gained from the generating
functional
\[
G[\phi]={\rm e}^{- S_G[\phi]} = \langle {\cal
T}\exp\left[\frac{i}{e}\int_{\cal C} dt I(t)
\phi(t)\right]\rangle\, ,
\]
where $I(t)$ is the current operator and ${\cal T}$ the
time ordering operator along the Kadanoff-Baym contour
${\cal C}$. Time correlation functions of arbitrary order
of the current are determined from functional derivatives
of $G[\phi]$, in particular, the average current
\[
C_1(t)=\langle I(t)\rangle= i e\left. \partial
S_G[\phi]/\partial\phi(t)\right|_{\phi=0}
\]
and the current auto-correlation function
\[
C_2(t,t')=\langle I(t)\, I(t')\rangle=e^2\left.\partial^2
S_G[\phi]/\partial\phi(t)\partial\phi(t')\right|_{\phi=0}\,
.
\]
Higher order functional derivatives give the cumulants
related to non-Gaussian current fluctuations
\[
C_n(t_1,\ldots,t_n)=-(-i e)^n \left.\partial^n
S_G[\phi]/\partial
\phi(t_1)\cdots\partial\phi(t_n)\right|_{\phi=0}\, .
\]
We remark that the functional $S_G[\phi]$ carries the full
frequency dependence of all current cumulants and not just
their time averaged zero frequency values usually studied
in the field of full counting statistics \cite{lll}.

By way of example let us consider an Ohmic resistor of
resistance $R$ in thermal equilibrium at inverse
temperature $\beta$. Then, the functional $S_G[\phi]\equiv
S_R[\phi]$ takes the well-known form
\[
S_R[\phi]=\frac{1}{2} \frac{h}{e^2 R}\,\int_{\cal
C}dt\int_{\cal C}dt'\, \alpha(t-t')\ \phi(t)\, \phi(t')
\]
where
\begin{equation}
\alpha(t)=\frac{\pi}{2\, (\hbar\beta)^2\ {\rm sinh}^2(\pi
t/\hbar\beta)}\, .
 \label{eq6}
\end{equation}
The quadratic form reflects the Gaussian nature of the
current fluctuations in this case which implies that all
cumulants except for $C_2$ vanish. On the other hand, for
a tunnel junction with many transmission channels, where
each channel has a small transmission coefficient $T_i$
leading to the dimensionless conductance
 $g_T=h/(4\pi\, e^2 R_T) =\pi\sum_i T_i$, where
$R_T$ is the tunneling resistance, one has \cite{zaikin}
\begin{equation}
S_T[\phi]=-4 g_T \int\limits_{\cal C}\! dt\int\limits_{\cal
C}\! dt'\, \alpha(t-t')
\sin^2\left[\frac{\phi(t)-\phi(t')}{2}\right]. \label{eq7}
\end{equation}
Here, the periodicity in $\phi$ reflects the discreteness
of the transferred charges associated with non-Gaussian
current fluctuations.

To gain information on the noise of the conductor, it may
be placed in parallel to a current biased JJ as depicted
in the circuit diagram of Fig.~\ref{fig1}. For a bias
current $I_b$ below the critical current $I_c$, the JJ is
in its zero voltage state and the bias current flows as a
supercurrent entirely through the JJ branch of the
circuit. Consequently, no heating occurs in the conductor
and the total system can easily be kept at low
temperatures, where the decay of the zero voltage state
occurs through MQT. The rate of this process depends with
exponential sensitivity on the current fluctuations of the
conductor so that the JJ acts as a noise detector.

The MQT rate $\Gamma$ can be calculated in the standard way
\cite{cl,ol} from the imaginary part of the free energy
$F$, i.e.,
\[
\Gamma=\frac{2}{\hbar} \mbox{ Im}\{F\}\, ,
\]
where $F=-(1/\beta) \ln(Z)$ is related to the partition
function $Z={\rm Tr}\{{\rm e}^{-\beta H}\}$. In the path
integral representation one has
\[
Z=\int {\cal D}[\theta] {\rm e}^{-S[\theta]}\, ,
\]
which is a sum over all imaginary time paths with period
$\hbar\beta$ of the phase difference $\theta$ across the
JJ weighted by the dimensionless action
$S[\theta]=S_{JJ}[\theta]+S_G[\theta/2]$. Here
\begin{equation}
S_{JJ}[\theta]={1\over \hbar}\int_{0}^{\hbar\beta}d\tau
\left[\frac{1}{2}\varphi_r^2 C_J
\dot{\theta}(\tau)^2+U(\theta)\right]\, \label{jjaction}
\end{equation}
is the action of the bare JJ and $S_G$ is the generating
functional of current fluctuations of the conductor
introduced above. In Eq.~(\ref{jjaction})
$\varphi_r=\hbar/2e$ denotes the reduced flux quantum,
$C_J$ is the capacitance of the JJ, and the tilted
washboard potential $U(\theta)=-E_J [\cos(\theta)-s\,
\theta]$, where $E_J$ is the Josephson energy and
$s=I_b/I_c$. The factor of 2 in the argument of $S_G$
arises from the fact that the voltage across the conductor
equals the voltage $V_J=(\hbar/e)(\dot{\theta}/2)$ across
the JJ.

In the MQT regime the partition function of the isolated JJ
is dominated by the so-called bounce trajectory, an
extremal $\delta S[\theta]=0$ periodic path in the
inverted barrier potential. By approximating a
well-barrier segment of $U(\theta)$ around a well minimum
$\theta_m$ by a harmonic+cubic potential,
$V(\delta\theta)=(M\Omega^2/2)\, \delta\theta^2
(1-\delta\theta/\delta\theta_0)$ with
$\delta\theta=\theta-\theta_m$, one finds an analytic
solution in the limit of vanishing temperature, i.e.,
\[
\theta_B(\tau)=\frac{\delta\theta_0}{{\rm
cosh}^2(\Omega\tau/2)}\, .
\]
Here, $\Omega=\Omega(s)$ is the frequency for small
oscillations around the well bottom (plasma frequency) at
bias current $s$, $M=\varphi_r^2 C_J$, and
$\delta\theta_0$ denotes the exit point determined from
$U(\theta_m)=U(\theta_m+\delta\theta_0)$. The
corresponding MQT rate reads
\[
\Gamma_0=6 \sqrt{6 \Omega V_b/\hbar\pi}\
\exp\left(-\frac{36}{5}\frac{V_b}{\hbar\Omega}\right)
\]
where $V_b=(2 M\Omega^2/27)\,\delta\theta_0^2$ is the
barrier height .

Following the theory of the effect of an electromagnetic
environment on MQT \cite{cl}, the partition function can
now be calculated for arbitrary coupling between detector
and conductor based on a numerical scheme developed in
\cite{ol}. Analytical progress is made when the noise
generating element has a dimensionless conductance $g\ll
E_J/\hbar\Omega$ so that the influence of the noise on the
MQT rate can be calculated by expanding about the
unperturbed bounce which gives
\begin{equation}
\Gamma=\Gamma_0\ {\rm e}^{-S_G[\theta_B/2]}\, .
\label{mqtrate}
\end{equation}
The correction $S_G[\theta_B/2]$ is usually dominated by
the second cumulant $C_2$ (width) and the fourth cumulant
$C_4$ (sharpness). Note that this approximation still
contains the full dynamics of detector and noise source
since any approximation relying on a time scale
separation, as e.g.\ the adiabatic limit considered in
\cite{pek1}, is usually not applicable.

Now, in case of a tunnel junction \cite{remark} as noise
element one finds for $S_G[\theta_B/2]\equiv
S_T[\theta_B/2]$ from (\ref{eq6}) and (\ref{eq7})
\begin{equation}
S_T[\theta_B/2]=\frac{g_T}{4\pi}\int_0^\infty d\omega\,
\omega\, |\tilde{\rho}(\omega)|^2\label{correction}
\end{equation}
with
\[
\tilde{\rho}(\omega)=\int_{-\infty}^\infty d\tau\, {\rm
e}^{i\theta_B(\tau)/2}\ {\rm e}^{i\omega\tau}\, .
\]
By expanding the first exponential and performing the
Fourier transform for each power of $\theta_B(\tau)$
separately,  the relevant part
$\rho(\omega)=\tilde{\rho}(\omega)-2\pi\delta(\omega)$
reads
\begin{equation}
\rho(\omega)=\frac{\pi}{4}\frac{\omega}{{\rm
sinh}(\pi\omega/\Omega)}\sum_{k=1}^\infty
 \frac{(2
i\delta\theta_0)^k}{k!\, (2k-1)!}\,
\prod_{l=1}^{k-1}\left(\frac{\omega^2}{\Omega^2}+l^2\right).
\label{rhoexpl}
\end{equation}
This way, Eq.~(\ref{correction}) can be cast into
\begin{equation}
 S_T[\theta_B/2]=\frac{g_T}{4
\pi^3}\sum_{k,k'=1}^\infty\ I_{k,k'} \,
\delta\theta_0^{k+k'}\label{corr2}
\end{equation}
 with the coefficients
\[
I_{k,k'}=\frac{ (-1)^{(3k+k')/2}\, 2^{k+k'}\, }{k! k'!
(2k-1)! (2k'-1)!}\ A_{kk'}.
\]
Here,
\begin{eqnarray}
A_{kk'}&=&\int_0^\infty dy \frac{y^3 {\rm e}^y}{({\rm
e}^{y}-1)^2}
\left[\prod_{l=1}^{k-1}\left(\frac{y^2}{4 \pi^2}+l^2\right)\right]\nonumber\\
&&\times\left[\prod_{l=1}^{k'-1}\left(\frac{y^2}{4 \pi
^2}+l^2\right)\right] \nonumber
\end{eqnarray}
with $A_{kk'}=A_{k'k}$ so that $I_{k,k'}\neq 0$ only for
$k+k'$ even. This means that all odd cumulants of the
fluctuating current vanish according to a vanishing net
current $\langle I(t)\rangle=0$ through the conductor.
Specifically, one finds
\begin{eqnarray}
A_{11}&=&6\zeta(3)\nonumber\\
A_{22}&=&6\zeta(3)+\frac{5!\,\zeta(5)}{2\pi^2}
+\frac{7!\,\zeta(7)}{16\pi^4}\nonumber\\
A_{31}&=&24\zeta(3)+5 \frac{ 5!\,\zeta(5)}{4\pi^2}
+\frac{7!\,\zeta(7)}{16\pi^4}\, .\nonumber
\end{eqnarray}
The terms in the sum (\ref{corr2}) related to a
contribution of order $\delta\theta_0^{k+k'}$ determine
the impact of the $(k+k')$th-moment of the current
fluctuations of the tunnel junction onto the MQT process.
Since in Eq.~(\ref{rhoexpl}) the term of order
$\delta\theta_0^{k}$ contains contributions centered
around $\omega\approx 0,\Omega\ldots,k\Omega$, the
influence of the $(k+k')$th-moment results from mode
mixing between fluctuations with frequencies $ l\, \Omega$
and $l'\, \Omega$ where $l\leq k, l'\leq k'$.

\begin{figure}
\epsfxsize=8cm
\epsffile{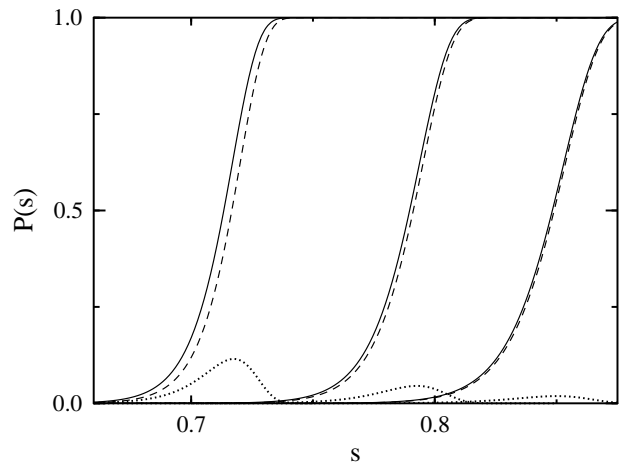} \caption[]{\label{fig2}Switching
probabilities out of the zero voltage state ("s-curve") of
a JJ in parallel to a tunnel junction (solid) and to an
ohmic resistor (dashed) with identical second cumulant for
various pulse lengths of the bias current, from left to
right: 1ms, 1$\mu$s, 10ns. Dotted lines below each pair of
s-curves display the differences of the corresponding
switching probabilities. Parameters are
$\sqrt{E_J/E_C}=10, g_T=2$, $\Omega(s=0)$=100 GHz.}
\end{figure}

In lowest order, $k+k'=2$, one gains from
Eq.~(\ref{corr2}) the Gaussian noise contribution
providing a correction to the bare MQT rate
\begin{equation}
\Gamma_{T}^{(2)}=\Gamma_{0}\, \exp\left[-\frac{6\zeta(3)\,
g_T}{\pi^3}\, \delta\theta_0(s)^2\right]\, .
\label{secondrate}
\end{equation}
Apparently, this reflects the well-known fact that Gaussian
noise leads to a {\em reduction} of the tunneling rate
\cite{cl}.

At order $\delta\theta_0^{4}$ the sum (\ref{corr2}) gives
three contributions, namely, $k=1, k'=3$ and $k=3$, $k'=1$
with $A_{13}=A_{31}$ as well as $k=2, k'=2$ with $A_{22}$.
This leads to
\begin{equation}
\Gamma_T^{(4)}=\Gamma_{T}^{(2)}\, \exp\left[\frac{4 \, g_T
}{\pi^3}\, (2A_{31}-A_{22})\, \delta\theta_0(s)^4\right]
\label{forthrate}
\end{equation}
so that the fourth order cumulant of the current noise
contains both, fluctuations that suppress tunneling
(related to $A_{22}$) and fluctuations that increase MQT
(related to $A_{31}$). Since $2A_{31}-A_{22}>0$, the total
impact of the fourth moment leads to an {\em enhancement}
of the MQT rate.

To obtain explicit results, one derives for the
harmonic+cubic potential $V(\delta\theta)$ an expression
for the amplitude $\delta\theta_0(s)$ of the bounce,
namely,
\[
\delta\theta_0(s)=3 \frac{\sqrt{1-s^2}}{s}\, .
\]
Accordingly, the barrier height scales with the
dimensionless current $s$ as $V_b(s)=(2E_J/3)
(1-s^2)^{3/2}/s^2$, and the plasma frequency reads
$\Omega(s)=[\sqrt{2 E_J E_C}/\hbar]\, (1-s^2)^{1/4}$ with
charging energy $E_C=2e^2/C$. Experimentally, in the
standard procedure \cite{jjmeasure} to measure the MQT
rate, a bias current pulse of height $I_b$ and duration
$t$ is adiabatically turned on and a voltage pulse is
detected when the JJ switches to its finite voltage state.
This procedure is performed a few thousand times to built
up switching histograms that determine the switching
probabilities
\[
P(s)=1-{\rm e}^{-\Gamma(s)\, t}\, .
\]
In Fig.~\ref{fig2} these so-called s-curves are shown for
various values of the duration of the current pulse $t$.
The parameters chosen are accessible in realistic
experiments. Apparently, for short pulses when the
switching occurs for values of the bias current close to
the critical current of the JJ, the effect of the
non-Gaussian noise fluctuations is completely suppressed
due to a decreasing amplitude $\delta\theta_0(s)$ of the
bounce. However, for longer pulses (or equivalently, for
shorter pulses in the tails towards lower $s$ values) the
fourth order cumulant leads to a substantial influence,
mostly dominated by a shift to smaller $s$-values compared
to an Ohmic resistor with identical Gaussian noise
contribution. As seen from Eq.~(\ref{forthrate}) this is
due to an effective decrease of the barrier height by the
non-Gaussian fluctuations of the tunnel junction. The
contrast in the switching probabilities between
 MQT in presence of purely Gaussian noise and in presence
 of a tunnel junction is larger than 10\% for values of the
 pulse height around $s=0.73$. This reflects the
 exponential sensitivity of the JJ in the MQT range even to
 weak non-Gaussian fluctuations.

 \begin{figure}
\vspace{0.5cm}
 \epsfxsize=8cm
\epsffile{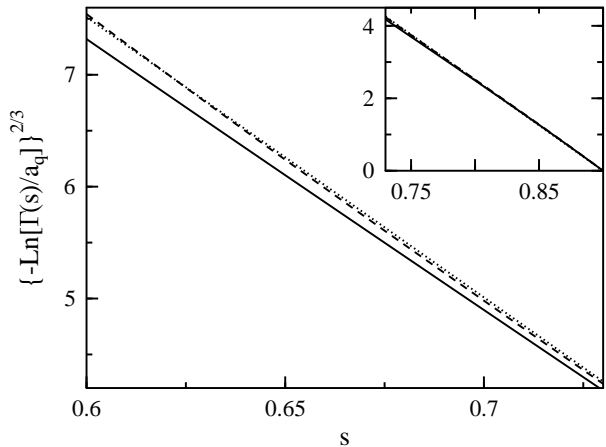} \caption[]{\label{fig3}
$B(s)=\{-{\rm ln}[\Gamma(s)/a]\}^{2/3}$ vs. the
dimensionless bias current $s$ for a tunnel junction
(solid) and an ohmic resistor (dashed) with identical
second cumulant. In the range of larger values for $s$
(inset) a straight line (dotted) is fitted to the data.
This reference line almost coincides with the data for
Gaussian fluctuations (dashed) while non-Gaussian noise
(solid) leads to a pronounced reduction of $B(s)$.
Parameters are the same as in Fig.~\ref{fig2}.}
\end{figure}

For the on-chip detection circuit proposed here, it may be
difficult to see the impact of the fourth order cumulant
directly in the s-curves since reference curves for purely
Gaussian noise and the same JJ are not available. In the
standard analysis of MQT data, the rate is obtained by
exploiting the $s$-dependence of the bounce action.
Namely, it turns out that the function
\[
B(s)=\{-{\rm ln}[\Gamma(s)/a_q]\}^{2/3}
\]
with the prefactor $a_q=6 \sqrt{6 \Omega V_b/\hbar\pi}$
gives rise to an essentially straight line, thus reflecting
the $s$-dependence of $V_b$. This holds true also for weak
purely Gaussian noise, but any non-Gaussian contributions
to the MQT rate can be expected to lead to deviations from
this scaling behavior. For this purpose, $s$-curves for
shorter pulses, i.e.\ larger $s$-values, where
non-Gaussian effects are absent, can be used to determine
the slope $b$ of $B(s)-B(s^*)=b (s-s^*)$ (reference point
$s^*$ close to 1). For longer pulses corresponding to
smaller values of the switching current, this straight
line $b (s-s^*)$ is then compared with the actual $B(s)$ as
depicted in Fig.~\ref{fig3}. Purely Gaussian noise shows
basically no deviations from the linear behavior over a
wide range of $s$-values, while closer inspection of the
analytical rate expression (\ref{secondrate}) reveals that
small deviations lead to a slight {\em increase} of the
actual $B(s)$ towards lower $s$. In contrast, non-Gaussian
noise displays substantial deviations already for much
larger values of $s$ and leads to a {\em decrease} of the
actual $B(s)$ compared to the reference line extracted for
$s$ close to 1. Thus, plotting $B(s)$ allows to directly
discriminate the impact of higher than second order
cumulants in the noise fluctuations of the conductor.
Further, by fitting $B(s)$ with Eq.\ (\ref{secondrate}),
the coefficient $2A_{31}-A_{22}$ related to $C_4$ in Eq.\
(\ref{forthrate}) can be extracted.

The formulation developed above in Eq.~(\ref{mqtrate}) is
completely general and applies to any nanoscale conductor
in parallel to a JJ. A systematic expansion of the action
$S_G[\theta_B/2]$ in powers of $(\theta_B/2)^n$ determines
the dynamical impact of the $n$th order cumulant $C_n$ onto
the MQT process.

To summarize we have proposed a nanoelectrical circuit
where a JJ placed in parallel to an arbitrary conductor
acts as detector for non-Gaussian current noise. Since no
net current flows through the noise source, heating
effects are suppressed and one obtains access to the even
order cumulants of the distribution function which are
notoriously difficult to detect. For experimentally
realistic parameters we have explicitly shown how the
fourth order cumulant of a tunnel junction can be
extracted.

Fruitful discussions with N.\ Birge, D.\ Esteve, T.\
Heikkil{\"a}, B.\ Huard, and H.\ Pothier
 are gratefully acknowledged. JA is a
Heisenberg fellow of the DFG.


\begin{thebibliography}{99}
\bibitem{remark} A finite capacitance $C_T$ of the tunnel junction
can easily be taken into account by replacing $C_J$ by
$C=C_J+C_T$.
\end{thebibliography}
\end{document}